\documentclass[useAMS,usenatbib]{mn2e}

\usepackage{graphicx}
\usepackage{amssymb}
\usepackage{amstext}
\usepackage{url}
\usepackage{amsmath}
\usepackage{setspace}
\usepackage{xtab}
\usepackage{textcomp}
\usepackage{booktabs,array}
\usepackage{epstopdf}
\usepackage{color}
\usepackage{colortbl}
\usepackage{tabularx}
\usepackage{gensymb}
\usepackage{subfigure}
\usepackage{placeins}
\usepackage{float}

\newcommand{\cd}{d\ensuremath{^{-1}}}
\newcommand{\kms}{\ensuremath{\textrm{km\,s}^{-1}}}
\newcommand{\vsini}{\ensuremath{v\sin i}}

\newcommand{\msun}{\ensuremath{\textrm{M}_{\odot}}}
\newcommand{\rsun}{\ensuremath{\textrm{R}_{\odot}}}

\title[Frequencies in the hybrid star HD\,49434]{The classification of frequencies in the $\gamma$ Doradus / $\delta$ Scuti hybrid star HD\,49434 \thanks{This paper
includes data taken at the Mount John University Observatory of the
University of Canterbury New Zealand, the McDonald
Observatory of the University of Texas USA, and the
European Southern Observatory at La Silla Chile.}}
\author[E. Brunsden, K. R. Pollard, P. L. Cottrell, K. Uytterhoeven, D. J. Wright, P. De Cat]
{E. Brunsden$^{1}$\thanks{E-mail: emily.brunsden@gmail.com}, K. R. Pollard$^{2}$, P. L. Cottrell$^{2}$, K. Uytterhoeven, D. J. Wright$^{3}$,
\and P. De Cat$^{4}$\\
$^{1}$Department of Physics, University of York, Heslington, York, YO10 5DD, UK\\ 
$^{2}$Department of Physics and Astronomy, University of Canterbury, Private Bag
4800, Christchurch, New Zealand\\
$^{3}$Department of Astrophysics, University of New South Wales, Sydney, NSW 2052
Australia\\
$^{4}$Royal Observatory of Belgium, Ringlaan 3, 1180 Brussel, Belgium}

\begin{document}

\date{ }

\pagerange{\pageref{firstpage}--\pageref{lastpage}} \pubyear{2014}

\maketitle

\label{firstpage}

\begin{abstract} Hybrid stars of the $\gamma$ Doradus and $\delta$ Scuti pulsation types have great potential for asteroseismic analysis to explore their interior structure. To achieve this, mode identifications of pulsational frequencies observed in the stars must be made, a task which is far from simple. In this work we begin the analysis by scrutinizing the frequencies found in the CoRoT photometric satellite measurements and ground-based high-resolution spectroscopy of the hybrid star HD\,49434. The results show almost no consistency between the frequencies found using the two techniques and no characteristic period spacings or couplings were identified in either dataset. The spectroscopic data additionally show no evidence for any long term (5 year) variation in the dominant frequency. The $31$ spectroscopic frequencies identified have standard deviation profiles suggesting multiple modes sharing ($l,m$) in the $\delta$ Scuti frequency region and several skewed modes sharing the same ($l,m$) in the $\gamma$~Doradus frequency region. In addition, there is a clear frequency in the $\gamma$~Doradus frequency region that appears to be unrelated to the others. We conclude HD\,49434 remains a $\delta$ Scuti/ $\gamma$ Doradus candidate hybrid star but more sophisticated models dealing with rotation are sought to obtain a clear picture of the pulsational behaviour of this star.

\end{abstract}


\section{Introduction}

The discovery of numerous candidate hybrid $\gamma$ Doradus/ $\delta$ Scuti stars by space telescopes, is proving to be a real problem for asteroseismology. Several $\gamma$ Doradus/$\delta$ Scuti hybrid stars have now been detected and studied from ground-based observations and further candidates are still being discovered with data from the MOST, CoRoT and \textit{Kepler} space telescopes. These stars all show frequencies, attributed to pulsation, in both the $\delta$ Scuti and $\gamma$ Doradus frequency ranges. Despite the impressive quality of the data, no mode identification has yet been published of such a hybrid star to confirm their nature. The detailed spectroscopic study of the candidate hybrid star HD\,49434 presents an opportunity to confirm the existence of hybrid stars and draw conclusions on their nature. 

Early \textit{Kepler} results show that $\gamma$ Doradus/$\delta$ Scuti hybrid stars may challenge current understanding of the classical instability
strip and the definitions of $\delta$ Scuti and $\gamma$ Doradus stars (\citeauthor{2010AN....331..989G} \citeyear{2010AN....331..989G}a, \citeauthor{2010ApJ...713L.192G} \citeyear{2010ApJ...713L.192G}b, \citeauthor{2011AandA...534A.125U} \citeyear{2011AandA...534A.125U}). Whilst many of the first hybrid stars
studied from space show hundreds of excited modes, other
detailed investigations of the frequencies show many of these hundreds are aliases, combinations or couplings of one another
\citep{2011MNRAS.414.1721B,2011AandA...525A..23C,2012AandA...540A.117C,2012arXiv1209.4836B}. Other studies have shown that granulation could be responsible for
many of these
frequencies,
especially those
previously attributed to high degree $l$ modes which suffer cancellation effects in photometry, even for high-precision space telescopes \citep{2010ApJ...711L..35K}.

Examination of the statistics of the class identifications is also proving to be problematic. The large numbers of hybrid stars found all over the $\delta$
Scuti and $\gamma$ Doradus regions of the Hertzsprung-Russell diagram are as yet unexplained
\citep{2010AN....331..989G,2011AandA...534A.125U,2011MNRAS.415.1691B} and are particularly difficult to explain in the hotter spectral-type A stars. These stars
are beyond the theoretical limit for driving of $\gamma$ Doradus-type pulsation and, indeed, any type of low-frequency pulsation. The alternate explanation,
starspots,
is also problematic as the lack of a sizeable convective shell for these hot stars inhibits starspot formation \citep{2011MNRAS.415.1691B}. To shed light on all
these questions arising from space photometry, more ground-based observations of stars, such as HD\,49434, are required to characterise the frequency spectra of
hybrid stars.

Historically, HD\,49434 (HR\,2514, HIP\,32617) was a remarkably understudied star given it has a visual magnitude of
5.74 \citep{2004AandA...425..683B}. Since its identification
as a $\gamma$ Doradus star by \citet{2002AandA...389..345B} however, much more effort has been put into several photometric and spectroscopic
studies. The significant recent studies of 
\citet{2008AandA...489.1213U} and \citet{2011AandA...525A..23C} have demonstrated that this star has frequencies in the $\gamma$ Doradus region ($0.3-3$~\cd) and in the $\delta$ Scuti region ($3-80$~\cd) and this star thus offers the first detailed view of a hybrid $\gamma$ Doradus/ $\delta$ Scuti star.

HD\,49434 has a spectral type of F1V and effective temperature of $7632~\pm~126$ K \citep{2006AandA...448..341G}. This places the star squarely in the expected overlap region of the p-mode $\delta$ Scuti pulsations and the g-mode $\gamma$ Doradus pulsations. Rotational velocity measurements show this star is a rapid rotator; the
most precise value is from \citet{2006AandA...448..341G} giving \vsini~$=85.4~\pm~6.6$~\kms. The same paper also contains a detailed abundance analysis, generally finding metallicities slightly higher than solar values. The abundances, in general,
agree
with previous spectroscopic work by \citet{2002AandA...389..345B}. The spectroscopic parameters from these works and those from two further
photometric studies by \citet{2001AandA...365..535L} and \citet{2005AJ....129.2461P} are given in Table~\ref{sum_param4}. The
radius of the star is estimated from \textsc{2mass} infra-red photometry to be $1.601~\pm~0.052$~\rsun\ \citep{2006AandA...450..735M}
and an estimate of the
stellar mass of $1.55~\pm~0.14$~\msun\ is given in \citet{2002AandA...389..345B}.

Analysis of pulsations in HD\,49434 began using ground-based spectroscopy in 1998 \citep{2002AandA...389..345B} when low-amplitude variations were observed in the
spectral-line profiles. Str\"omgren photometry, taken in 2001, confirmed the variability with excess power in the $1$~\cd\ to $5$~\cd\ region, but yielded no
identifiable
frequencies \citep{2002AandA...389..345B}. Further spectroscopy by \citet{2004AandA...417..189M} showed, again, low-amplitude variations in the blue wing of
the line profile. 

This star has been observed at Mt John University Observatory (\textsc{mjuo}) since February 2007 when a two-week observing run as part of a multi-site
campaign was
undertaken. The purpose of the campaign was to gather spectroscopic data simultaneous with observations from the CoRoT photometric
satellite. The results were published in \citet{2008AandA...489.1213U}. From the spectroscopic data, three $\gamma$ Doradus and five $\delta$ Scuti
frequencies were identified, leading to the classification of HD\,49434 as a hybrid pulsator. Only one of the $\gamma$ Doradus and none of the $\delta$ Scuti
frequencies were detected in the accompanying ground-based photometry from Sierra Nevada Observatory, Spain and San Pedro M\'artir Observatory, Mexico. The frequencies from the CoRoT data analysis are presented
in \citet{2011AandA...525A..23C}. The high-precision photometry revealed a staggering $1686$ significant frequencies in the five-month observing run with
$840$ of these identified as pulsation frequencies. The
vast number of frequencies found have led to further questions regarding the origins and stability of the stellar pulsations. 

\begin{table}\caption{Summary of stellar parameters of
 HD\,49434 derived from spectroscopic (first two entries) and photometric (latter two entries) studies. Figures in parentheses represent uncertainties of their respective values.}\label{sum_param4}
\begin{center}
\begin{tabular}{lcccc}
\toprule
Paper & $T$ & log$g$ & [Fe/H] & \vsini \\
 & (K) &  & & (\kms ) \\
\midrule
G\&M\footnotemark[1] & 7632 (126) & 4.43 (0.20) & 0.09 (0.07) & 85.4 (6.6)  \\ 
B\&al\footnotemark[2] & 7300 (200) & 4.14 (0.20) & -0.13 (0.14) & 84 (4)  \\ 
\\
P\&al\footnotemark[3] & 7250 (200) & 4.1 (0.2) & -0.1 (0.2) &   \\ 
L\&al\footnotemark[4] & 7240 (100) & 4.0 (0.4) & -0.1 (0.2) &   \\ 
\bottomrule
\end{tabular}
\end{center}
\footnotemark[1] \citet{2006AandA...448..341G}
\footnotemark[2]\citet{2002AandA...389..345B}
\footnotemark[3]\citet{2005AJ....129.2461P}
\footnotemark[4]\citet{2001AandA...365..535L}
\end{table}

This work expands on the analysis of the $689$ spectra in \citet{2008AandA...489.1213U} with the analysis of a further $1058$ multi-site spectra making a total of $1747$ observations.

\section{Observations and Data Treatment}\label{comp4}

\begin{figure}
\centering
   \includegraphics[width=0.45\textwidth]{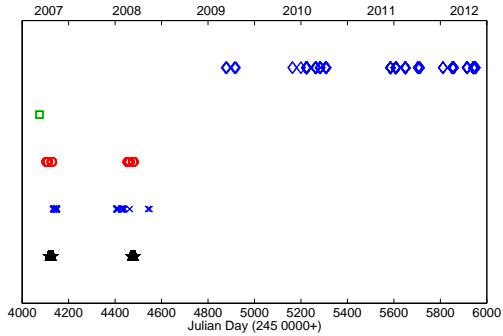}
\caption[Observation log of the four data sites for HD\,49434.]{Timing of spectra obtained from the four data sites; \textsc{hercules} data in blue diamonds and blue crosses, \textsc{foces} data in green squares, \textsc{feros} data in red circles and \textsc{sophie} data in black stars. The clusters of observations loosely match 2007 to 2012
observation seasons. The two \textsc{hercules} datasets
were reduced using different image processing software and are regarded as independent sites. The prior-published data in
\protect\citet{2008AandA...489.1213U} corresponds to the first cluster of observations in all datasets from the 2007 observing season around JD 245\,4100.
}\label{obs_multi}
\end{figure}

Observations of HD\,49434 have continued at \textsc{mjuo} from the initial run in 2007 until January 2012, contributing to a large, long-term dataset to confirm and refine these frequencies. These data were combined with line
profiles computed in \citet{2008AandA...489.1213U} with additional spectra from \textsc{feros} and \textsc{sophie} taken in 2008. A summary of the times
and dates of observations of the individual sites is presented in Table~\ref{sum_obs4}. These observations are plotted in Figure~\ref{obs_multi} to show the temporal distribution of the data.

Spectra from \textsc{mjuo} in New Zealand were collected using the $1$~m McLellan telescope with the fibre-fed High Efficiency and Resolution Canterbury University Large \'Echelle Spectrograph (\textsc{hercules}) with a resolving power of R $= 50000$ operating over a range of $3800~$\AA\ to $8000~$\AA\ \citep{2002ExA....13...59H}. The data collected from \textsc{mjuo} was divided into two time periods. The oldest dataset, which contained observations taken in 2007 and 2008, was analysed
as part of the multi-site analysis of \citet{2008AandA...489.1213U}. Since then, observations have been taken from 2009 to 2012 which were analysed as a
single-site dataset. This dataset had a total of $381$ observations. Spectra were reduced using a \textsc{matlab} pipeline written by Dr. Duncan Wright.
This pipeline performs the basic steps of flat fielding from white-lamp observations, calculating a dispersion solution from thorium-lamp observations and outputting the data
into a two dimensional format. 

\textsc{feros} is now installed on the Max-Planck-Gesellschaft / \textsc{eso} $2.2$\,m telescope and has a resolving power of R $= 48000$ \citep{1999Msngr..95....8K}. The
spectral range covered is $3500$~\AA\ to $9200$~\AA. Reductions were done using a pipeline written for \textsc{feros} in \textsc{midas} \citep{wtf}.

The $1.93$\,m telescope at the Observatoire de Haute Provence, France, can observe with \textsc{sophie} (Spectrographe pour l'Observation des
Ph\'enom\`enes des Int\'erieurs stellaires et des Exoplan\`etes) \citep{2008SPIE.7014E..17P}. The wavelength range is $3870$~\AA\ to $6940$~\AA\ with a resolving power of R
$= 70000$ (high-resolution mode). A software package adapted from the High Accuracy Radial velocity Planet Searcher (\textsc{harps}) spectrograph was used to
reduce the spectra.

\begin{table*}\caption{Summary of spectroscopic observations of HD\,49434 from 2006 to 2012.}\label{sum_obs4}
\begin{center}
\begin{tabular}{lccccc}
\toprule
Spectrograph & Observatory & Tel. & Observation range & $\Delta$T (d) & $\#$ Obs. \\
\midrule
\textsc{foces}  & Calar Alter & 2.2 m & Dec 2006 & 1 & 47\\
\textsc{feros}  & La Silla & 2.2 m & Jan 2007-Jan 2008  & 376 & 404\\
\textsc{sophie} & Haute Provence & 1.93 m & Jan 2007-Jan 2008  & 374 & 629\\
\textsc{hercules} (old)  & {\footnotesize \textsc{mjuo}} & 1.0 m & Feb 2007-Mar 2008  & 416 & 286\\
\textsc{hercules} (new)  & {\footnotesize \textsc{mjuo}} & 1.0 m & Feb 2009-Jan 2012  & 1076 & 381\\
\bottomrule
\end{tabular}
\end{center}
\end{table*}

Each spectrum was cross-correlated using the delta-function method (\citeauthor{NewEntry3}, \citeyear{NewEntry3}, building on \citeauthor{2007CoAst.150..135W}, \citeyear{2007CoAst.150..135W}) to
create a single representative line-profile from approximately $4000$ lines in the full spectrum. To check the consistency of methods between this paper and that of \citet{2008AandA...489.1213U}, a selection of the data taken as part of the
2007 multi-site spectroscopic campaign was analysed in \textsc{famias}. These data were provided as line profiles (as discussed in
\citeauthor{2008AandA...489.1213U}, \citeyear{2008AandA...489.1213U}) which were computed using a Least-Squares Deconvolution (\textsc{lsd}) method
\citep{1997MNRAS.291..658D,1999A&AS..134..149D}. This method differs
slightly to the $\delta$-function cross-correlation method used for the more recent \textsc{hercules} data, but produces very similar line profiles.

\begin{figure}
\centering
   \includegraphics[width=0.45\textwidth]{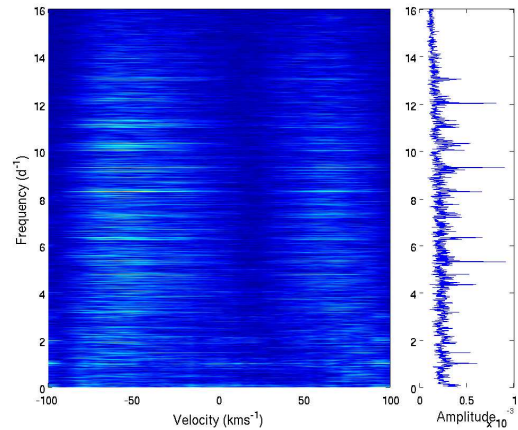}
\caption{The
2D periodogram (left) showing the frequencies found in the \textsc{LSD} line profile across each pixel (in velocity measurements) (lighter colours for stronger peaks). The mean periodogram over the 48 pixels is plotted on the right.}\label{twod_all}
\end{figure}

Time-series analysis was performed using \textsc{famias} \citep{2008CoAst.157..387Z}, applied as in \citet{2006AandA...455..235Z}. This entailed an analysis of
the variations in the representative line profiles which identified the frequencies present. For examples of $\gamma$ Doradus stars analysed using these techniques see \cite{2012MNRAS.422.3535B}a, \cite{2012arXiv1209.6081B}b or \cite{2014PASA...31...25D}. A periodogram of the amplitude of the variation at each pixel in the line profile (in velocity measurements) for each frequency is shown in Figure \ref{twod_all} as an example of the pixel-by-pixel (\textsc{pbp}) technique. Frequency analysis of the moments (radial velocity, width and skewness) relies on the average symmetry of the line profile over a pulsation period. This is not
the case for HD\,49434 and the strong asymmetry of the variation of the line profile (see Figure \ref{stddel} and Figure \ref{stdgam}) results in non-sinusoidal behaviour of these
measurements, limiting the detection of the frequencies. Further comments on the origins and repercussions of the skew line profile is
included in the discussion (Section~\ref{discskew}). 

A comparison between the results of the two data sets showed the pixel-by-pixel analysis to be nearly identical to the results of the line
profile variations in \citet{2008AandA...489.1213U}. These results show excellent consistency between the Discrete Fourier Transform method employed in \textsc{famias} compared with the Intensity Period Search \citep{1997AandA...317..723T} method used in \citet{2008AandA...489.1213U}. The consistency between the results means we can
merge the line profiles into a single extensive observational dataset and indicates no significant change to the results from the calculation of the line profile
using the $\delta$-function cross-correlation or an \textsc{lsd} technique.

For $\gamma$ Doradus and $\delta$ Scuti stars, the standard deviation and phase profiles show the motion of the pulsation though the spectral line. These profiles vary depending on the mode, frequency and stellar parameters which can be determined using the mode identification tools in \textsc{famias}. In general, different modes have distinctive shapes in the line profiles and smooth phase profiles across the line with jumps linked to stationary points of the line profile. Examples of typical $\gamma$ Doradus and $\delta$ Scuti line, standard deviation and phase profiles can be seen in \citet{2012MNRAS.422.3535B} and \citet{2014AandA...570A..33S}, respectively. 

Spectroscopic mode identification relies on the rotational broadening of the line profile to return a use number of pixels in the line profile. To date stars from $\vsini = 9$~\kms \citep{2001ApJ...553..814A} to $\vsini = 101$ \kms \citep{meow} have been analysed successfully although it is noted that \textsc{famias} does have a mode-dependent rotation limit as deviations of the star from spherical symmetry due to rotation are ignored. 

%%%%%%%%%%%%%%%%%%%%%%%%%%%%%%%%%%%%%%%%%%%%%%%%%%%%%%%%%%%%%%%%%%%%%%%%%%%%%%%%%%%%%%%%%%%%%%%%%%%%%%%%%%%%%%%%%%%%%%%%%%%%%%%%%%%%%%%%%%%%%%%%%%%%%

\section{Frequency Analysis}\label{full4}

\begin{table}\caption[Frequencies found in the \textsc{pbp}
analysis of the full HD\,49434 dataset.]{Frequencies found in the \textsc{pbp}
analysis of the full HD\,49434 dataset. A $\dagger$ represents frequencies discarded as aliases as described in the text. Values in column 2 in bold
type were also identified in the first $100$ detected frequencies of
CoRoT.
Frequencies underlined indicate those in common with the identification of
\protect\citet{2008AandA...489.1213U}, with (*) being found only using the moment method. The \textsc{pbp} amplitudes are measured in normalised intensity units. Column 5 indicates group members of \textsc{dsl}, 
\textsc{gdl}, \textsc{irr} and 
\textsc{spc} as described in the text.}\label{pixelall}
\begin{center}
\begin{tabular}{lcccc}
\\
\hline
ID & Freq. (\cd) & Amplitude & Rel Amp. & Class. \\
\hline
$	f_{1}	$	&	\underline{\textbf{	9.3071	}}	&	0.29	&	1.00	&	DSL	\\
$	f_{2}	$	&	\underline{	5.3298	}	&	0.28	&	0.97	&	DSL	\\
$	f_{3}	$	&	\underline{	12.0330	}	&	0.21	&	0.72	&	DSL	\\
$	f_{4}	\dagger$	&		0.0003		&	0.14	&	0.48	&	IRR	\\
$	f_{5}	$	&	\textbf{	3.1779	}	&	0.15	&	0.52	&	DSL	\\
$	f_{6}	$	&		4.8073		&	0.18	&	0.62	&	DSL	\\
$	f_{7}	$	&	\underline{\textbf{	1.4855	}}	&	0.14	&	0.48	&	GDL	\\
$	f_{8}	$	&	\textbf{	6.6781	}	&	0.1	&	0.34	&	DSL	\\
$	f_{9}	$	&	\textbf{	4.7432	}	&	0.13	&	0.45	&	DSL	\\
$	f_{10}	$	&		9.1385		&	0.11	&	0.38	&	DSL	\\
$	f_{11}	$	&	\underline{\textbf{	5.5843	}}*	&	0.11	&	0.38	&	DSL	\\
$	f_{12}	$	&		5.7879		&	0.11	&	0.38	&	DSL	\\
$	f_{13}	\dagger$	&		0.9993		&	0.12	&	0.41	&	IRR	\\
$	f_{14}	$	&	\textbf{	1.8744	}	&	0.11	&	0.38	&	GDL	\\
$	f_{15}	$	&		7.7628		&	0.09	&	0.31	&	DSL	\\
$	f_{16}	$	&		12.5488		&	0.08	&	0.28	&	IRR	\\
$	f_{17}	\dagger$	&	\textbf{	0.2237	}	&	0.07	&	0.24	&	GDL	\\
$	f_{18}	$	&		6.9081		&	0.08	&	0.28	&	DSL	\\
$	f_{19}	\dagger$	&		1.0008		&	0.09	&	0.31	&	IRR	\\
$	f_{20}	\dagger$	&	\textbf{	0.4077	}	&	0.07	&	0.24	&	GDL	\\
$	f_{21}	\dagger$	&	\textbf{	0.2493	}	&	0.07	&	0.24	&	GDL	\\
$	f_{22}	$	&		6.6292		&	0.06	&	0.21	&	DSL	\\
$	f_{23}	\dagger$	&		0.0003		&	0.14	&	0.48	&	IRR	\\
$	f_{24}	$	&		8.2185		&	0.05	&	0.17	&	DSL	\\
$	f_{25}	$	&		7.4212		&	0.07	&	0.24	&	DSL	\\
$	f_{26}	\dagger$	&	\textbf{	0.0556}		&	0.08	&	0.28	&	IRR	\\
$	f_{27}	$	&		9.3978		&	0.06	&	0.21	&	DSL	\\
$	f_{28}	$	&		12.0326		&	0.08	&	0.28	&	IRR	\\
$	f_{29}	\dagger$	&		0.0805		&	0.06	&	0.21	&	IRR	\\
$	f_{30}	$	&		2.0513		&	0.06	&	0.21	&	IRR	\\
$	f_{31}	$	&	\textbf{	2.5382	}	&	0.07	&	0.24	&	SPC	\\
\hline
\end{tabular}
\end{center}
\end{table}

\begin{figure}
\centering
  \includegraphics[width=0.45\textwidth, trim=1cm 0.2cm 0.1cm 0.1cm, clip=true]{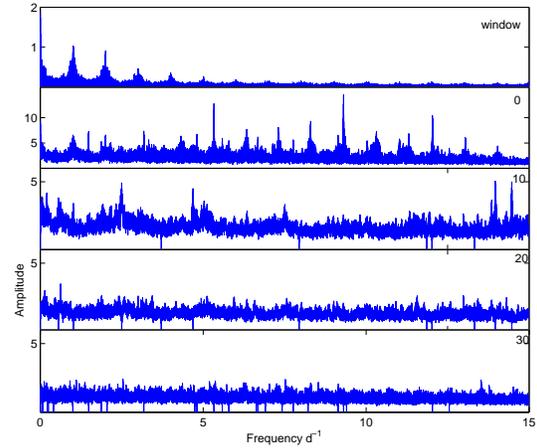}
   \caption[Fourier spectra showing the original spectrum and the prewhitening stages.]{Fourier spectra showing the spectral window, original spectrum and the
prewhitening
stages after the removal of $10, 20$ and $30$
frequencies. Note the change in scale of the y-axis. Fourier spectra are scaled by a factor of $10\,000$.}\label{stdpbpfourier}
\end{figure}

The frequencies were analysed solely using the pixel-by-pixel method due to the severe asymmetry of the line profile. In total $31$ frequencies ($f_1$ to $f_{31}$) were identified from the peaks in the Fourier spectra that were above the detection limit as defined in \textsc{famias}. A full list is shown
in Table~\ref{pixelall} and the Fourier spectra at selected prewhitening stages are shown in Figure~\ref{stdpbpfourier}. Formal uncertainties on the frequencies
range from $\pm~0.00003$~\cd\ for $f_4$ to $\pm~0.0002$~\cd\ for $f_{31}$, but due to the low amplitude of the pulsations and the numerous observational
uncertainties
inherent
in the data, the frequencies are reported with a conservative error of $\pm 0.001$~\cd. 

Upon inspection, six of these frequencies ($f_4$, $f_{13}$, $f_{19}$, $f_{23}$, $f_{26}$ and $f_{29}$) were flagged as possible aliases due to their
proximity to
$0$~\cd\ or $1$~\cd, but were kept for the subsequent analysis of the aliases standard deviation profiles. Artificially increased amplitudes in the Fourier spectrum
are possible due to higher noise levels particularly in the $0$ to $1$~\cd region, which can push a noise peak higher than a real frequency peak.

\subsection{Frequency Aliases and Combinations}\label{8alias}

The large number of frequencies found in the multi-site spectroscopic data warranted the computation and comparison of frequency aliases and combinations in an
automated way. For aliasing this focused the search for $f_a \pm p = f_b \pm 0.002$ for integer values of $p$ up to $p = 12$. Only the match $f_3 \approx f_{28}$ was identified.

Combinations of frequencies were tested for $p*f_a \pm q = r*f_b \pm 0.002$~\cd\ for integer values of $p, q$ and $r$ up to $12$ including $q=0$. Several
identifications were discovered but on closer examination these had high values of two or more of $p, q$ and $r$ (e.g. $3f_{1}~\approx~5f_{11}$) and are
unlikely
to arise without the additional presence of simpler combinations. As such they are regarded as purely coincidental. It is also unlikely that combinations not
involving the highest amplitude frequencies would arise without seeing higher amplitude combinations from $f_1$ to $f_3$.

\subsection{Comparison with CoRoT Results}\label{8compcorot}

The results of the dedicated CoRoT study of more than $331\,000$ single-passband photometric observations, taken over nearly $140$ days from October 2007 to
March
2008, are presented in \citet{2011AandA...525A..23C}. They find a staggering $1686$ significant frequency peaks and propose $840$ frequencies intrinsic to
HD\,49434. The frequencies and their properties are available in electronic format as supplementary information to the paper. The uninterrupted high quality of
the dataset allows for high
precision
of the detected frequencies and these can be compared with the spectroscopic results to understand the variability of this star. In the following analyses, the
full $1686$ frequencies are considered.

The ten frequencies with the highest amplitude from the CoRoT analysis were first compared with the spectroscopic results to see if any frequencies
matched. The frequencies are given in Table~\ref{corot10}. Only the
frequencies $f_{c2}$, $f_{c3}$, $f_{c6}$ and $f_{c7}$ were found in the spectroscopic analysis. The frequencies $f_{c1}$ to $f_{c10}$ were then imposed on the
spectroscopic data to observe any potential signal and characterise the standard deviation and phase profiles. None of the ten frequencies have clear line profile and phase variations easily attributable to pulsations. 

\begin{table}\caption[First ten frequencies detected with the CoRoT satellite.]{The ten highest-amplitude frequencies of $840$ detected with the CoRoT satellite from
\protect\citet{2011AandA...525A..23C}.
These frequencies were tested for occurrence in the \textsc{pbp} data.}\label{corot10}
\begin{center}
\begin{tabular}{ccccc}
\toprule
ID & Frequency  & Amp.  & Amp. & Signif. \\
 & (\cd) & (mmag) & (rel to 1) &  \\
\midrule
$f_{c1}$	&	1.73463954 (092)	&	2.4599	&	1.00	&	15558.3	\\
$f_{c2}$	&	2.53808114 (142)	&	1.5891	&	0.65	&	8284.4	\\
$f_{c3}$	&	0.22356263 (145)	&	1.5534	&	0.63	&	8962.6	\\
$f_{c4}$	&	2.25335725 (145)	&	1.5526	&	0.63	&	10201.6	\\
$f_{c5}$	&	0.40754452 (173)	&	1.3045	&	0.53	&	8416.4	\\
$f_{c6}$	&	0.24907883 (215) 	&	1.0522	&	0.43	&	6203	\\
$f_{c7}$	&	0.05537893 (221)	&	1.0213	&	0.42	&	6276.8	\\
$f_{c8}$	&	1.54418626 (236)	&	0.9578	&	0.39	&	6148.1	\\
$f_{c9}$	&	2.37947392 (243)	&	0.9298	&	0.38	&	6339	\\
$f_{c10}$	&	3.97738341 (251)	&	0.8991	&	0.37	&	6493.2	\\
\bottomrule
\end{tabular}
\end{center}
\end{table}

When the first $100$ frequencies of the CoRoT findings were compared to the \textsc{pbp} frequencies, $12$ of the $31$ \textsc{pbp} frequencies were found to
match to within $\pm
0.003$~\cd. These matches are listed in Table~\ref{pixelcomp}. It is surprising how poorly the frequencies detected in each method align, especially the
high-amplitude frequencies. 

\begin{table}\caption[Frequencies in common between the CoRoT photometry and spectroscopy.]{Frequencies in common between the
CoRoT photometry \protect\citep{2011AandA...525A..23C} and the spectroscopic analysis presented in this work.}\label{pixelcomp}
\begin{center}
\begin{tabular}{ccc|ccc}
\toprule
\multicolumn{3}{c}{CoRoT} & \multicolumn{3}{c}{\textsc{pbp}} \\
Freq. & ID & Rel. &  Freq &ID & Rel. \\
(\cd) &  &  Amp. &  (\cd) & &  Amp. \\
\midrule
0.055	&	7	&			0.40	&	0.056	&	$f_{26}$	&			0.3	\\
0.224	&	4	&			0.58	&	0.224	&	$f_{17}$	&			0.3	\\
0.249	&	14	&			0.40	&	0.249	&	$f_{21}$	&			0.3	\\
0.408	&	5	&			0.54	&	0.408	&	$f_{20}$	&			0.3	\\
1.485	&	30	&			0.09	&	1.485	&	$f_{7}$	&			0.5	\\
2.538	&	6	&			0.53	&	2.538	&	$f_{31}$	&			0.2	\\
2.871	&	33	&			0.08	&	1.874	&	$f_{14}$	&			0.3	\\
3.180	&	72	&			0.04	&	3.178	&	$f_{5}$	&			0.6	\\
4.585	&	32	&			0.08	&	5.584	&	$f_{11}$	&			0.4	\\
5.192	&	36	&			0.08	&	5.193	&	$f_{35}$	&			0.2	\\
6.678	&	52	&			0.05	&	6.678	&	$f_{8}$	&			0.3	\\
9.307	&	68	&			0.04	&	9.307	&	$f_{1}$	&			1	\\
\bottomrule
\end{tabular}
\end{center}
\end{table}

\subsection{Classification of Frequencies}

\begin{figure}
\centering
\includegraphics[width=0.45\textwidth,]{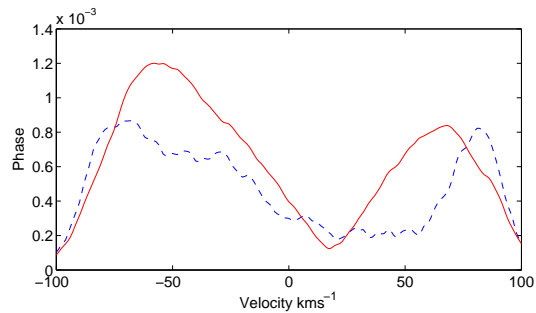}
\caption{Average standard deviation plots of the ``$\delta$-Scuti
like'' (\textsc{dsl}) frequencies (red, solid lines) and the ``$\gamma$-Doradus like'' (\textsc{gdl}) frequencies (blue, dashed lines).}\label{together}
\end{figure}

A least-squares fit was performed on the representative line profiles of the $1747$ observations using the profiles of the $31$ frequencies identified. The standard deviation profiles of each frequency fit were compared. Upon inspection, the profiles could be seen to fit into one of four classes. The mean profiles of the first two classes are plotted together in Figure \ref{together} for direct comparison.

\begin{figure}
\centering
\includegraphics[width=0.45\textwidth,]{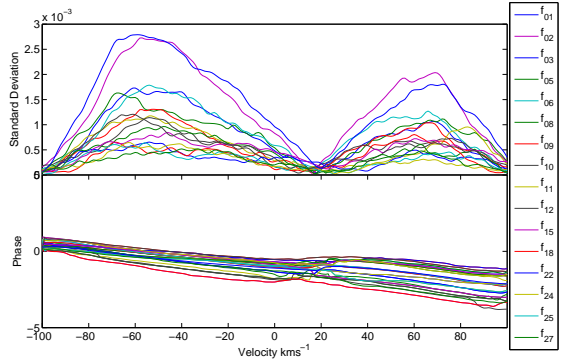}
\caption{Standard deviation plots of the ``$\delta$-Scuti
like'' (\textsc{dsl}) frequencies. }\label{stddel}
\end{figure}

The first group, with the majority of the frequencies, has a two-bump structure with peaks terminating at the edges
of the profile. The centres of the standard deviation and phase profiles are shifted to the right and are centred near $18$~\kms, an effect not seen in other $\gamma$-Doradus and $\delta$-Scuti stars. Nineteen of the
frequencies fitted this
description and they are plotted in Figure~\ref{stddel}. Noting the frequencies were all in the $\delta$ Scuti range, this group was named
``$\delta$-Scuti like'' (\textsc{dsl}).

\begin{figure}
\centering
\includegraphics[width=0.45\textwidth,]{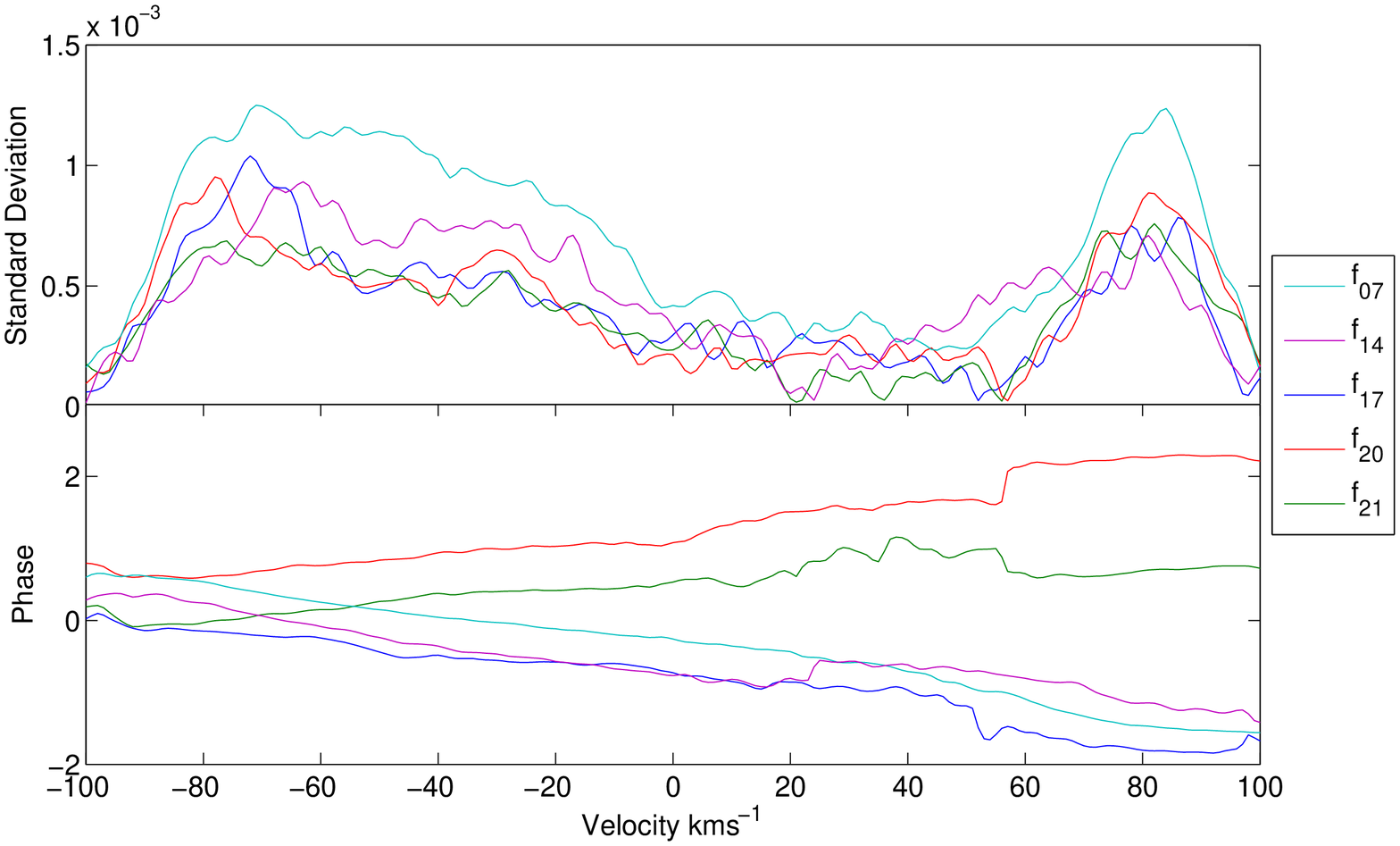}
\caption{Standard deviation plots of the ``$\gamma$-Doradus like'' (\textsc{gdl}) frequencies. }\label{stdgam}
\end{figure}

The second-identified group contained five of the frequencies and since these were found to be exclusively in the $\gamma$ Doradus
range this set was labelled ``$\gamma$-Doradus like'' (\textsc{gdl}). The standard deviations and phases are plotted in Figure~\ref{stdgam}. Note the
normalised phases of
two of the frequencies can be inverted to match those of the other three frequencies. 

\begin{figure}
\centering
\includegraphics[width=0.45\textwidth,]{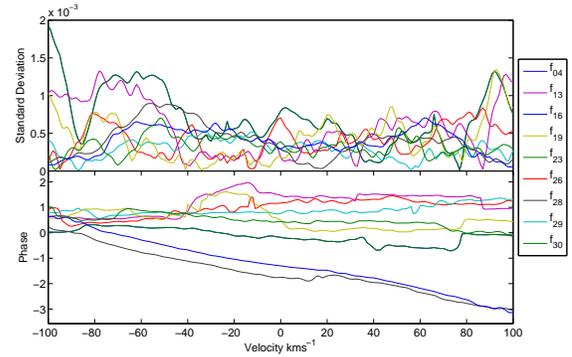}
\caption{Standard deviation plots of the ``irregular'' (\textsc{irr}) frequencies. }\label{stdirr}
\end{figure}

The third group of misshapen standard deviation profiles was named `` Irregular '' (\textsc{irr}) and individual profiles are plotted in Figure~\ref{stdirr}
with nine frequencies
of this type
found. Most show only small changes in phase. Many of these profiles have frequencies at zero or one cycle-per-day as described above and most are likely to be
artefacts of the data sampling. The two
exceptions are $f_{16}$ and $f_{28}$. The latter is almost certainly a residual from the removal of the strong $f_{3}$ frequency and the former could be an
alias or combination frequency, or a real frequency undistinguished from the noise. This may also be the case for some others in this group.

\begin{figure}
\centering
\includegraphics[width=0.45\textwidth,]{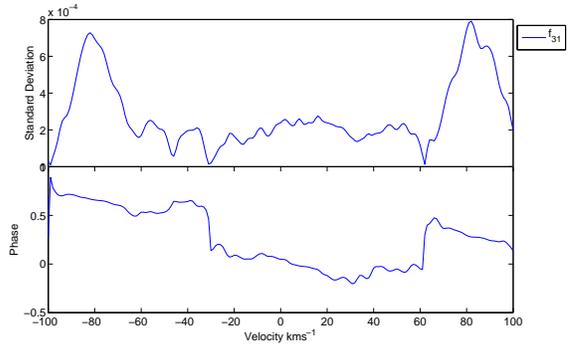}
\caption{Standard deviation plots of the one ``special case'' (\textsc{spc}) frequency. }\label{stdsp}
\end{figure}

The final group contained just one interesting frequency. The profile of $f_{31}$ had a shape similar to other identified profiles in fast-rotating $\gamma$
Doradus stars with narrow variation peaks in the wings and no variation in the line centre. This variation is matched by the changes in the phase profile. This
frequency is plotted in Figure~\ref{stdsp} and is labelled as a ``special case'' (\textsc{spc}).

\begin{figure}
\centering
\subfigure[Frequency $f_1$, a \textsc{dsl} frequency.]{  
\includegraphics[width=0.38\textwidth,trim=0.4cm 0.4cm 1cm 0.1cm, clip=true]{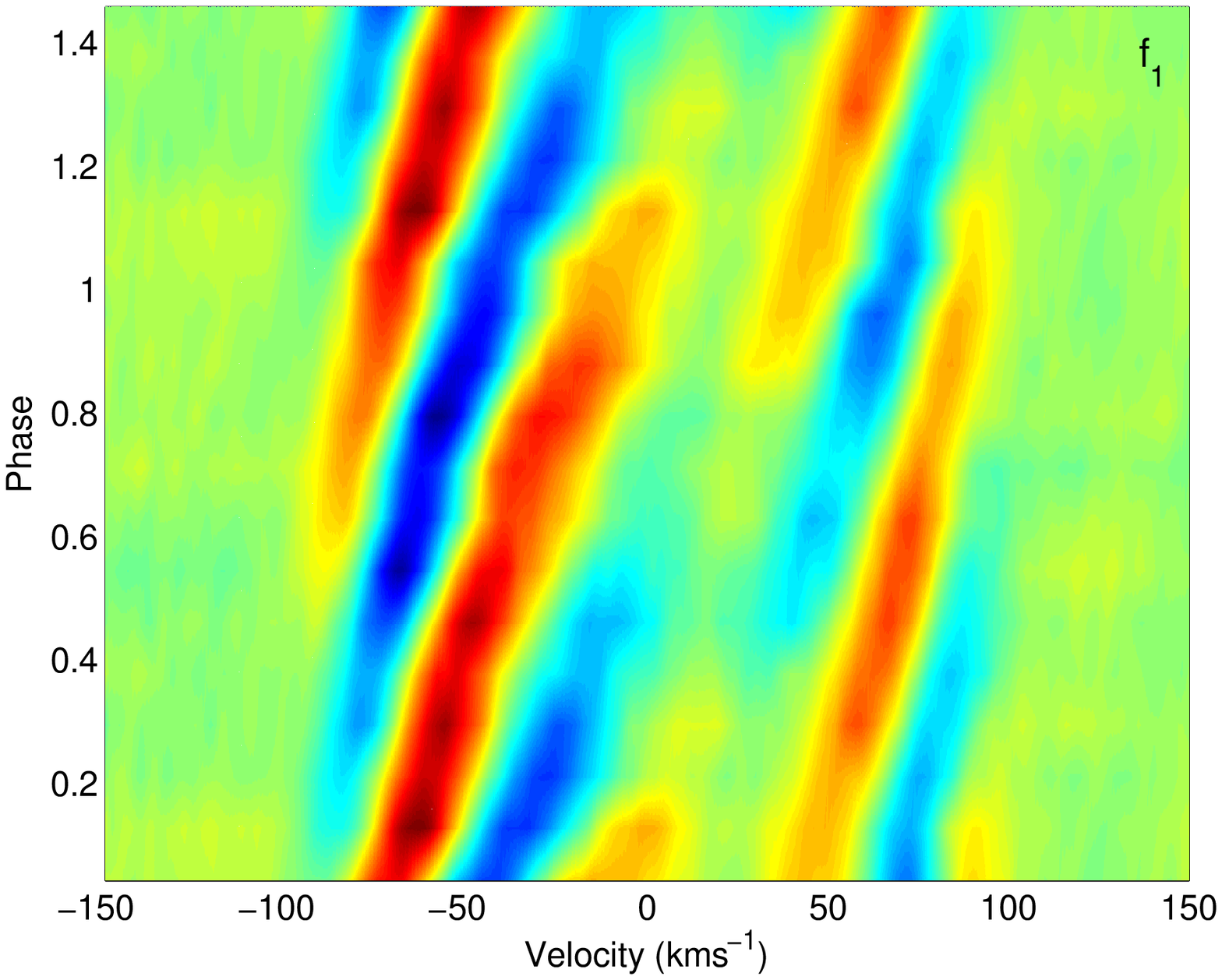}
}
\subfigure[Frequency $f_7$, a \textsc{gdl} frequency.]{  
\includegraphics[width=0.38\textwidth,trim=0.4cm 0.4cm 1cm 0.1cm, clip=true]{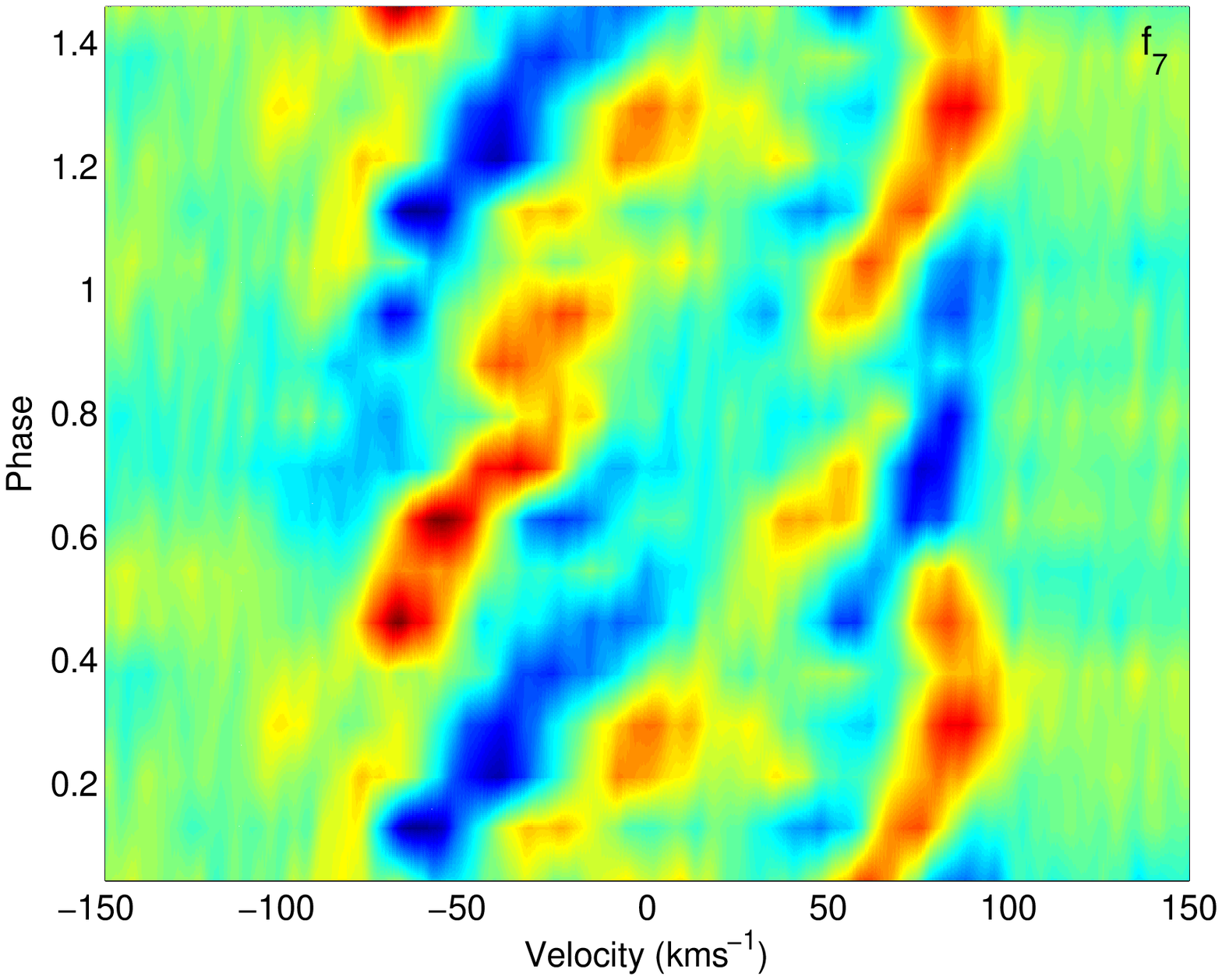}
}
\subfigure[Frequency $f_4$, an \textsc{irr} frequency.]{  
\includegraphics[width=0.38\textwidth,trim=0.4cm 0.4cm 1cm 0.1cm, clip=true]{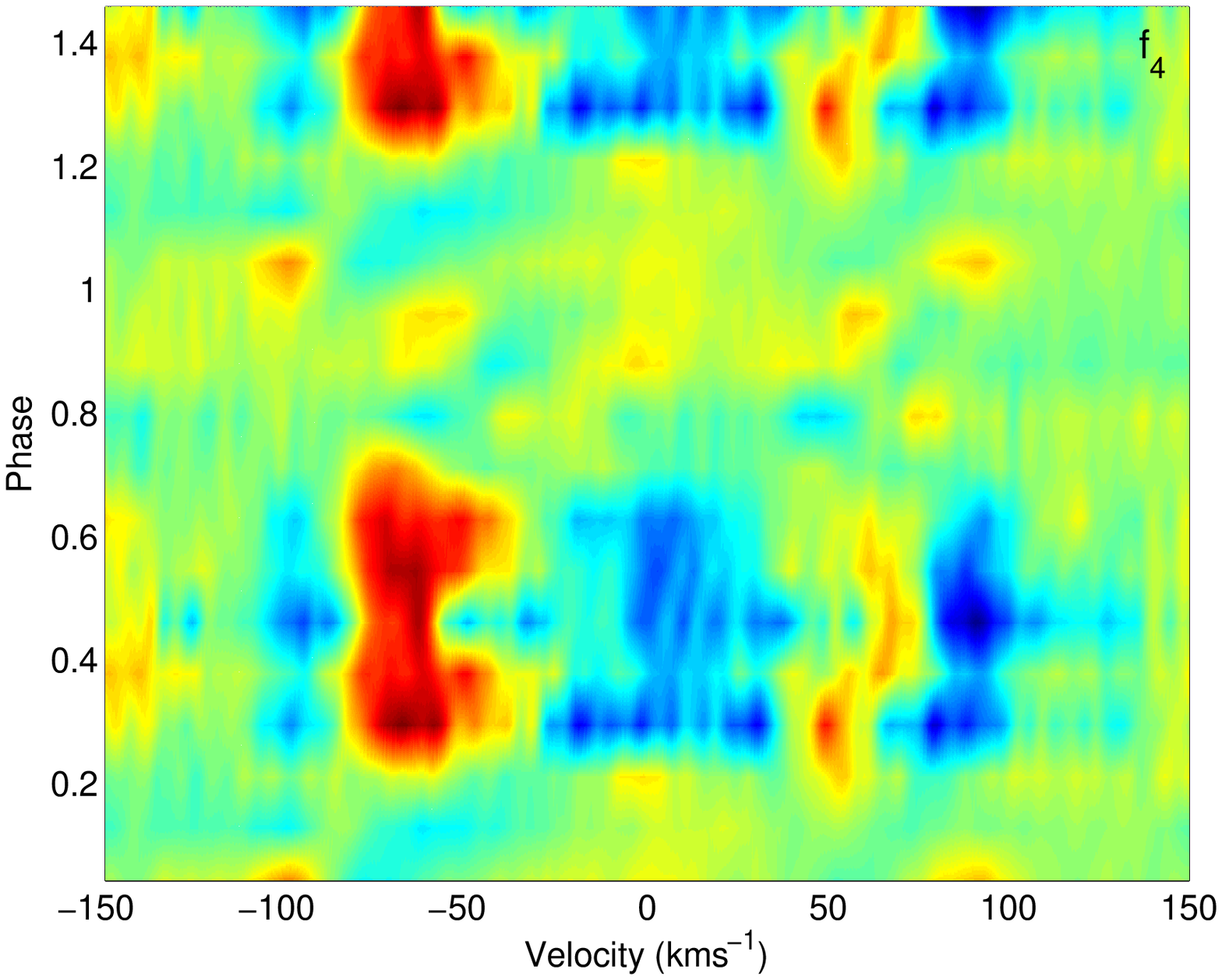}
}
\subfigure[Frequency $f_{31}$, the \textsc{spc} frequency.]{  
\includegraphics[width=0.38\textwidth,trim=0.4cm 0.4cm 1cm 0.1cm, clip=true]{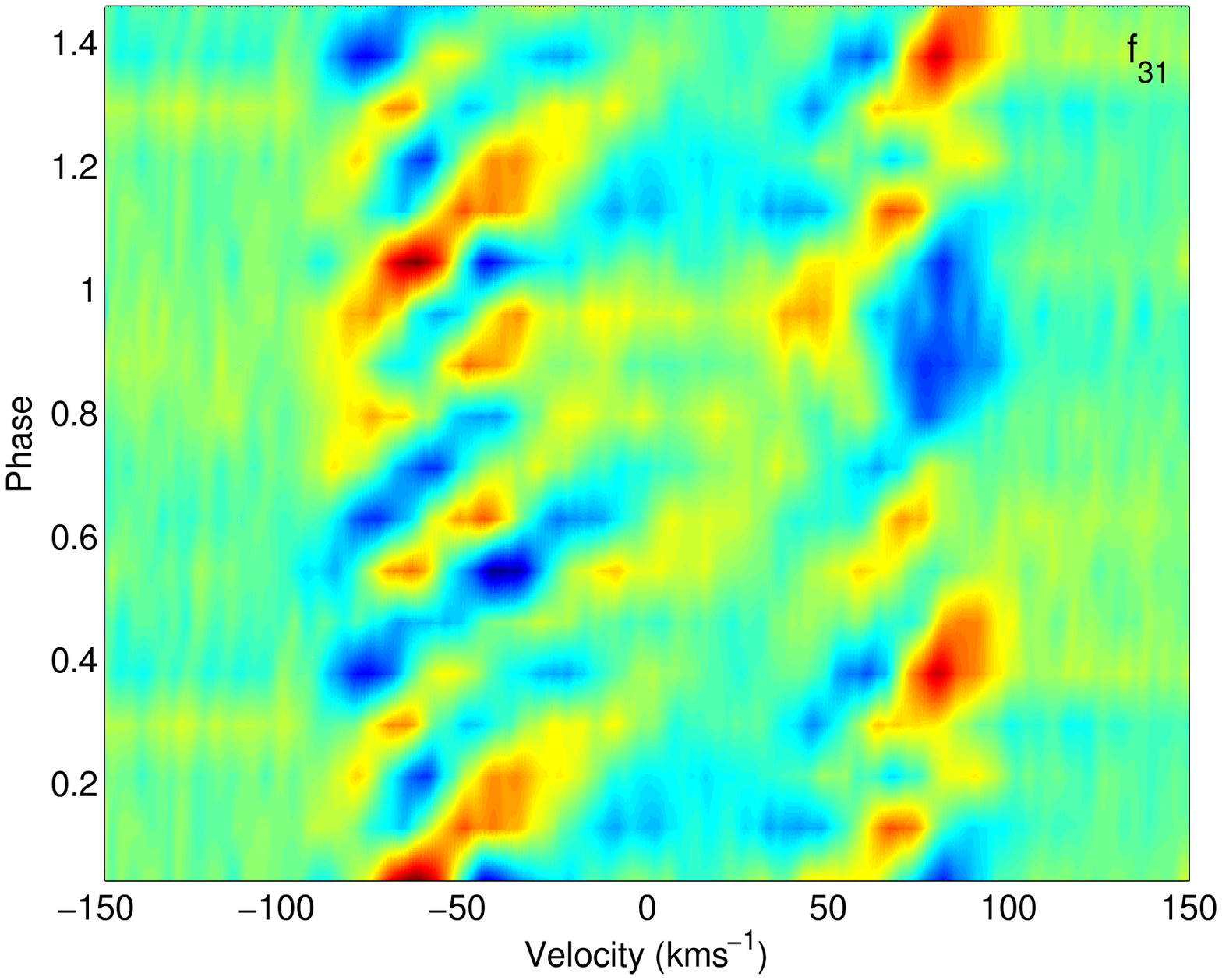}
}
\caption{Example phased residual line-profile plots of frequencies in the four classes.}\label{exampha8}
\end{figure}

To further investigate these groups, the residuals from the subtraction of the mean line profiles were phased over each of the $31$ identified frequencies in
Table~\ref{pixelall}. The results of the four groups are represented by the plots in Figure
\ref{exampha8}. These plots provide information on the stability of the frequency. Stellar frequencies are expected to show a smooth `braided rope' structure as the pulsation moves through the line profile. Different modes will also show different inclinations and structure through the pulsation phase. As with the standard deviation profiles, the plots fall into one of three categories. The first group
includes those plots with a clear structure with narrow `braids', mostly in two regions of the line profile. These plots generally correspond to
the $\delta$-Scuti like (\textsc{dsl}) frequencies. Most of the $\gamma$-Doradus like (\textsc{gdl}) frequencies show a similar `braided rope' shape with fewer
braids with a more gentle phase slope. Frequencies with no clear pulsation spreading across the entire phase were classed as irregular (\textsc{irr}) and one
frequency does not fit the above classifications, the special case (\textsc{spc}).

Based on the standard deviation profiles, phased residual profiles and the proximity of many to integer values, the frequencies in the \textsc{irr} group were
classified as non-stellar frequencies, likely arising from data sampling aliases. The frequency $f_{28}$ was also put in this group as it is the same as $f_3$ within the detection limit. The remaining
frequencies are regarded as candidate stellar frequencies which, together with the non-stellar frequencies, are further scrutinised in the following section.

The shape of the standard deviation profiles of the
\textsc{dsl} group could be fitted by a mode identification, but the central shift and uneven amplitude of the standard deviation profiles are not easily
accounted for. The homogeneity of the shapes of all the frequencies suggests that they are all representations of the same excited mode (couplings or splittings), or that there are many
excitations of modes with the same ($l,m$) and different ($n$) . The shape of the \textsc{gdl} standard deviation profiles is not typical of a pulsation and suggests that they may
be
artefact frequencies, possibly of some rotational effect. This topic is discussed further in Section~\ref{hybr}. The special case frequency, $f_{31}$, on the
other hand appears to be a real mode typical of a $\gamma$ Doradus pulsation. It is not clear why this mode shows such clean variation with low amplitude. The
shape of the standard deviation and phase profiles are recognisable in an individual least-squares fit, with only mild distortions, which rules out the
frequency being a
random residual from the interaction of other frequencies.

\subsection{Frequency Results}

The above analyses of all the frequencies detected in Table~\ref{pixelall} led to $9$ frequencies being discarded and $22$ retained as probable stellar
frequencies. Discarded frequencies are identified in Table~\ref{pixelall} by a $\dagger$.

% \begin{table}\caption[Stellar frequencies and their classifications for HD\,49434.]{Stellar frequencies and their classifications for HD\,49434. Values
% in column 2 and column 5 in bold type were also identified in the first $100$ detected frequencies of CoRoT.
% Frequencies underlined indicate those in common with the identification of
% \protect\citet{2008AandA...489.1213U}, with (*) being found only using the moment method.}\label{finalfreq8}
% \begin{center}
%   \begin{tabular}{ccc|ccc}
%  \toprule
% ID & freq. (\cd)& class. & ID & freq. (\cd) &  class.\\
% \midrule
% $	f_{1}	$	&	\underline{\textbf{	9.307	}}	&	DSL	&	$	f_{14}	$	&	\textbf{	1.874	}
% &	GDL	\\
% $	f_{2}	$	&	\underline{	5.330	}	&	DSL	&	$	f_{15}	$	&		7.763		&	DSL
% \\
% $	f_{3}	$	&	\underline{	12.033	}	&	DSL	&	$	f_{16}	$	&		12.549		&	DSL
% \\
% $	f_{5}	$	&	\textbf{	3.178	}	&	DSL	&	$	f_{18}	$	&		6.908		&	DSL
% \\
% $	f_{6}	$	&		4.807		&	DSL	&	$	f_{22}	$	&		6.629		&	DSL	\\
% $	f_{7}	$	&	\underline{\textbf{	1.485	}}	&	GDL	&	$	f_{24}	$	&		8.219		&
% DSL	\\
% $	f_{8}	$	&	\textbf{	6.678	}	&	DSL	&	$	f_{25}	$	&		7.421		&	DSL
% \\
% $	f_{9}	$	&	\textbf{	4.743	}	&	DSL	&	$	f_{27}	$	&		9.398		&	DSL
% \\
% $	f_{10}	$	&		9.138		&	DSL	&	$	f_{28}	$	&		12.033		&	DSL	\\
% $	f_{11}	$	&	\underline{\textbf{	5.584	}}*	&	DSL	&	$	f_{31}	$	&	\textbf{	2.538	}
% &	SPC	\\
% $	f_{12}	$	&		5.788		&	DSL	&				&				&		\\
% \bottomrule
% \end{tabular}
% \end{center}
% \end{table}

The detection of very different frequencies in photometry and spectroscopy of $\gamma$ Doradus
stars has been previously documented \citep[e.g. HD\,40745 in][]{2011MNRAS.415.2977M} but this was attributed to high noise levels in the spectroscopic
data. This hypothesis has not
been able to be excluded until the present work. The physical reason for the discrepancy is normally thought to be the low sensitivity of photometric techniques to
high degree modes due to
cancellation effects. This reasoning fails with the explanation of $f_{1}$ and $f_{2}$ because, as discussed in Section~\ref{full4}, the frequencies appear to
have the same standard deviation and phase profiles and therefore are likely to originate from modes with the same ($l$,$m$), or be otherwise linked. A full mode identification of the frequencies found is required to confirm this, the results of which will follow in a later publication.

%%%%%%%%%%%%%%%%%%%%%%%%%%%%%%%%%%%%%%%%%%%%%%%%%%%%%%%%%%%%%%%%%%%%%%%%%%%%%%%%%%%%%%%%%%%%%%%%%%%%%%%%%%%%%%%%%%%%%%%%%%%%%%%%%%%%%%%%%%%%%%%%%%%%%
\section{Probing the Hybrid Nature}\label{hybr}

The presence of both $\gamma$ Doradus and $\delta$ Scuti range frequencies suggests that HD\,49434 may be a hybrid. This
classification was made by \citet{2008AandA...489.1213U} based on the frequencies found and the proximity of the star to the intersection of the
pulsation groups in the Hertzsprung-Russell (\textsc{hr}) diagram \citep[see Figure 4 from][]{2002AandA...389..345B}. The shapes of the line profiles of the individual frequencies in Section~\ref{full4} suggests physical differences between the two groups of frequencies.
Additionally, two recent papers \citep{2012AandA...540A.117C,2012arXiv1209.4836B} have found links between the low-frequency and high-frequency modes of hybrid
stars
in space photometry. In this section the links between the two frequency groups are tested and the hybrid nature of this star is examined.

\subsection{The $\gamma$ Doradus/$\delta$ Scuti Frequency Domains}

Models of $\gamma$ Doradus/$\delta$ Scuti hybrid stars predict a gap in the frequency spectrum between the two types of
pulsation \citep{2005A&A...435..927D,2010AN....331..989G}. \citet{2012AandA...540A.117C} found the CoRoT hybrid star {105733033} had two
distinctive
frequency domains for the $\gamma$ Doradus and $\delta$ Scuti
frequencies identified. The same effect has also been observed in the \textit{Kepler} hybrid \textsc{kic}~{8054146} \citep{2012arXiv1209.4836B}. The
domains of
the
frequencies identified in HD\,49434 found using different methods do not show a clear separation between
the two frequency regions as in \citet{2012AandA...540A.117C}. This absence of domains could be due to a difference in rotation rate of the stars. \citet{2012AandA...540A.117C}
suggest the domains are
distinguishable due to the low rotation rate of the CoRoT star, implying the observed frequency spectrum is similar to the intrinsic spectrum of the co-rotating
frequencies of the star, but no measurement of the rotational velocity of the star has yet been made. If this were the case, it would therefore be unlikely the
same is true for HD\,49434, as the minimum equatorial rotational velocity is around $84~$\kms. This high rotational velocity would shift frequencies from the
co-moving frame according to rotational perturbation theory, and current perturbative methods (such as those used in \textsc{famias}) are not directly applicable
to stars with such high rotational
velocities to allow for an estimation of this \citep{2010A&A...518A..30B}. If this were the case, it is not clear why \textsc{kic}~{8054146}, with a
very high
rotation rate, does not
show the same phenomenon. 

\subsection{Radial Modes}

Further results from the \citet{2012AandA...540A.117C} study of CoRoT~{105733033} showed the majority ($187$ of $246$) of the $\delta$ Scuti
frequencies were couplings of the $\gamma$ Doradus frequencies and the dominant frequency, identified as the fundamental radial mode ($F$). The coupling
followed the pattern $p*F \pm f_{\gamma Dor}$ for $p= 1,2,3$ with amplitudes four times smaller than the original $\gamma$ Doradus frequencies ($f_{\gamma
Dor}$). This was also seen in \citet{2011MNRAS.414.1721B}, and further analysed in \citet{2011CoAst.162...62G}, for the $\delta$ Scuti star \textsc{kic}~{9700322}. Here the low frequency was
identified as the difference between two dominant radial modes. The physical explanation offered by \citet{2012AandA...540A.117C} for this effect is the
trapping of g-modes in the interior of the
star which may induce thermodynamic perturbations in the convective envelope where the p-modes originate. Many $\delta$ Scuti stars follow this model of a
dominant radial mode, so it could be expected for a hybrid star to display the same behaviour.

The dominant mode in the spectroscopic analysis of this star ($f_1~=~9.307$~\cd) is a $\delta$ Scuti frequency which could potentially be the radial fundamental
mode (usually A to F spectral-type stars have fundamental radial mode frequencies between $8$~\cd\ and $24$~\cd). The standard deviation profile of the
frequency
is a two-bump shape typical of the radial modes
\citep[see for example][]{2006AandA...455..235Z,2007A&A...471..237Z}. The phase changes of HD\,49434, however, are much smoother than the sharp ``jumps''
observed in the above examples. If $f_1$ was found to be the fundamental radial mode then the other \textsc{dsl} frequencies would be also interpreted as radial
modes. This cannot be the case as the low-frequency modes such as $f_5$ could not also be radial modes. Better mode-identification of the primary frequency is
required for a conclusive result. The spacings between this candidate
radial
mode and the other $\delta$ Scuti frequencies were tested for a correlation with the $\gamma$ Doradus frequencies for $p*F \pm q*f_{a} = f_b \pm 0.002$~\cd\ for
integer vales of $p$ and $q$ from one to five. No matches were found other than for $3F \approx 5f_{11}$ as found in Section~\ref{8alias}. 

\subsection{Characteristic Spacings}

In the asymptotic regime, frequencies with the same g-mode identification are predicted to be related by characteristic
period spacings
between frequencies of consecutive degree \citep{1980ApJS...43..469T}. The study of \citet{2012AandA...540A.117C} found $24$ of the $180$
identified $\gamma$ Doradus frequencies to have an asymptotic period spacing. Similarly, p-mode oscillations have a characteristic frequency spacing for modes
with sequential $n$. Given the possibility of both p- and g- modes being present in HD\,49434, both frequency and period spacings were tested for
sequencing. This search also could identify any equidistant frequencies arising from rotational splitting of the frequencies.

\begin{figure}
\centering
\subfigure[Frequency spacings in the spectroscopic frequencies.]{  
\includegraphics[width=0.45\textwidth]{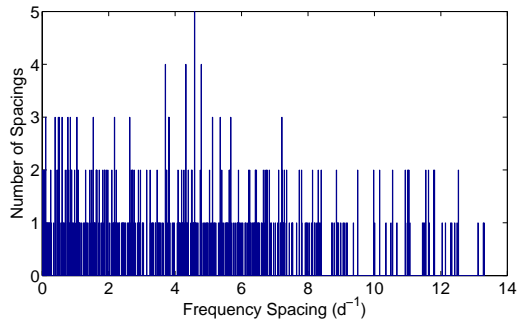}\label{specfhist1}
}
\subfigure[Period spacings in the spectroscopic frequencies.]{  
\includegraphics[width=0.45\textwidth]{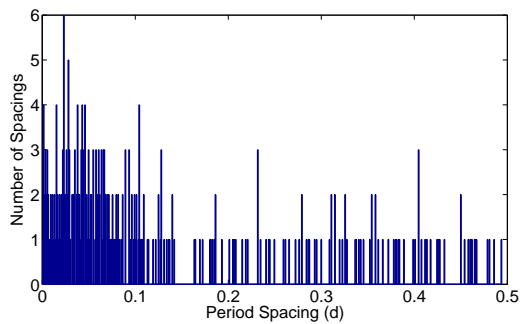}\label{specphist}
}
\subfigure[Frequency spacings in the CoRoT frequencies.]{  
\includegraphics[width=0.45\textwidth]{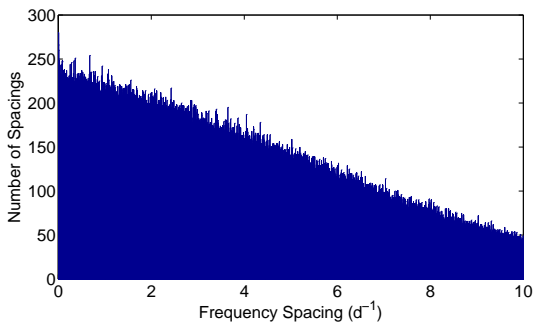}\label{corotfhist}
}
\subfigure[Period spacings in the CoRoT frequencies.]{  
\includegraphics[width=0.45\textwidth]{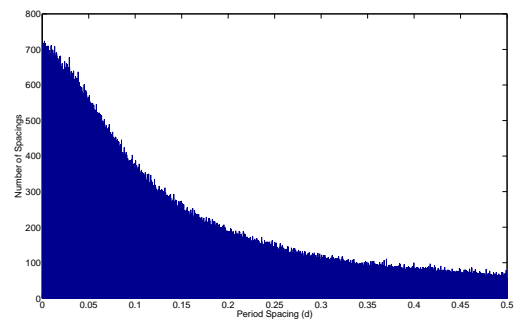}\label{corotphist}
}
\caption[Frequency and Period spacings found in spectroscopy and with CoRoT.]{Frequency and
Period spacings between all frequencies found in the spectroscopic analysis (Table~\ref{pixelall}) and the full CoRoT frequency identification set
\cite{2011AandA...525A..23C}.}\label{histspace}
\end{figure}

As with the aliases and combinations, this was tested
using an automated procedure by computing all possible spacings for the frequencies and corresponding periods identified in Table~\ref{pixelall} and
binning them using a resolution of $0.002$~\cd. The resulting histogram is shown as Figure~\ref{specfhist1} for the frequency spacing and Figure~\ref{specphist}
for the spectroscopic results. This was repeated for the CoRoT full frequency identification and selected frequency subset in Figure~\ref{corotfhist} and Figure
\ref{corotphist}.

A possible repeated frequency spacing at $4.585~$\cd\ and a possible repeated period spacing at $0.0235~$d were observed in the spectroscopic data. The CoRoT
data do not show
any single frequency or period spacing that is favoured, so it is likely the spectroscopic identification is coincidental, particularly as it does not involve
the strongest $\delta$ Scuti or $\gamma$ Doradus frequencies. If the $\delta$-Scuti range frequencies arise from one or more radial modes then it is not
expected any rotational splitting will be seen, unlike CoRoT 102918586 \citep{2010arXiv1004.1525M}. Additionally, the very high value of \vsini\ for this star
may mean it is a
very rapid rotator and splits may become asymmetric (splitting values vary by up to $0.008$~\cd in the above CoRoT star) and indistinguishable from other
frequency peaks. 

The above computation of the frequency spacings also allowed a comparison with the results of \citet{2012AandA...540A.117C} and
\citet{2012arXiv1209.4836B}, where frequency spacings from the dominant $\delta$ Scuti frequencies were found to match those of the high-amplitude $\gamma$
Doradus frequencies. A search for frequency combinations in the spectroscopic data revealed only a possible link between $f_7$, $f_8$
and $f_{35}$. Extended searches showed no other links between the strong frequencies. A general search for combinations in the CoRoT data reveals many candidate
groups of frequencies due to the large number of inputs. A more constrained search of the first $100$ identified frequencies for spacings, including one or more
frequencies above $4~$\cd\ with links to one of the $10$ highest-amplitude frequencies, found $10$ frequency groups with limits of $\pm 0.002~$\cd. None of
these combinations included any other of the $30$ highest-amplitude frequencies, so it is unlikely that the frequencies in HD\,49434 are showing the same
effects as the above hybrid stars.

\subsection{Frequency and Amplitude Variations}

Recent high-precision studies of $\gamma$ Doradus/$\delta$ Scuti hybrid stars \citep[e.g.][]{2012AandA...540A.117C,2012arXiv1209.4836B} have identified
variations
in the photometric amplitudes of observed frequencies over time, on a scale of months to years. The same phenomenon was tested using the spectroscopy of HD\,49434. The data
were
broken into observing year blocks, generally comprising of the southern-hemisphere summer observing season. The blocks correspond to the groups depicted in Figure~\ref{obs_multi}.
The years 2007 to 2012, excepting 2009, were tested as there were insufficient observations in the 2009 season for an independent analysis. The Fourier
spectrum
for each dataset was calculated and the highest amplitude frequency extracted. This was the equivalent of $f_{p1}$ for all but the 2010 dataset. The full
results are given in Table~\ref{ampvar}. It is not clear why $f_{p1}$ is not recovered in the 2010 dataset, but it may be due to unfortunate data sampling or
lower quality data. Despite this, the four distinct time periods are 
sufficient for an indication of any amplitude variability. The table shows the variation of the best fit amplitude over the five years of data to be a maximum
of $0.00037$ or 25\%. The frequencies themselves show no variation within the detection limits ($\pm 0.001$~\cd). The
amplitudes do appear to vary by up to 25\%, but it is not clear that this is not an effect of the different size and data quality encompassed in each set. All
but
the most extreme
amplitude variation would be extremely difficult to detect in the best ground-based spectroscopic campaigns and Blazhko-like variability (non-symmetric changes
in the maximum and minimum amplitudes), as seen in \citet{2012arXiv1209.4836B}, is beyond the precision of current spectroscopic methods. It is thus
still possible
that HD\,49434 has variations in the amplitudes of its pulsations, but this cannot be confirmed in this study.

\begin{table}\caption{Variation of the frequency and amplitude of $f_{p1}$ over time. The \textsc{pbp} amplitudes are measured in normalised intensity units and the final column is the least-squares fit amplitude of the frequency.}\label{ampvar}
\begin{center}
\begin{tabular}{cccccc}
\toprule
Year & \# sites & \# obs. & $f_{1}$ & Amp & Amp$_{fit}$\\
\midrule
2007 & 4 & 695 & 9.30707 & 0.00166 & 0.33286 \\
2008 & 4 & 671 & 9.30660 & 0.00141 & 0.28360 \\
2010 & 1 & 120 & \multicolumn{3}{c}{not recovered} \\
2011 & 1 & 157 & 9.30825 & 0.00129 & 0.26084 \\
2012 & 1 & 95 & 9.30785 & 0.00138 & 0.27827 \\
\bottomrule
\end{tabular}
\end{center}
\end{table}

%%%%%%%%%%%%%%%%%%%%%%%%%%%%%%%%%%%%%%%%%%%%%%%%%%%%%%%%%%%%%%%%%%%%%%%%%%%%%%%%%%%%%%%%%%%%%%%%%%%%%%%%%%%%%%%%%%%%%%%%%%%%%%%%%%%%%%%%%%%%%%%%%%%%%
\section{Discussion}\label{disc4}

The analysis of some of the best-available spectroscopic and photometric datasets on a $\gamma$ Doradus, or $\gamma$ Doradus hybrid star, has clearly
highlighted
the complexities and gaps in current understanding of these pulsators. This discussion focuses on three areas where ground has been broken in this study, yet
a full explanation is still lacking.

\subsection{Standard Deviation Profile}\label{discskew}

Skewed line profiles and slight asymmetries in the standard deviation profile have been observed in several other $\gamma$ Doradus stars, e.g. \textsc{hd}$\,${40745} \citep{2011MNRAS.415.2977M} and \textsc{hd}$\,${189631} \citep{meow}. This has usually been observed in the blue wing of the line
profiles. It is not yet clear whether this effect is intrinsic to the star or if it is an artefact of the line profile creation. Currently, two further
refinements to line profile generation are being developed to form part of the reduction package. These including the inclusion of a least-squares deconvolution to the cross-correlation technique and improvements to the weighting of the $\delta$-function amplitudes for individual lines. These may improve the signal in the line profiles in future.

This is the first $\gamma$ Doradus star that exhibits a severe shift in velocity space between the mean line profile and the standard deviation profile. The cause of
this relative shift is not known, yet it hampers significantly the mode identification. Both rotation and non-adiabatic surface effects could be proposed as the mechanism for the shift but more examples are required for a solid analysis. It follows that future modelling routines should be capable of fitting
this shift, and therefore shedding light on its physical origin.

Any rapidly-rotating star must be carefully analysed with the knowledge that there are not yet available pulsation models which include a full description of stellar
rotation. It is therefore likely that the inclination and rotation could skew line profiles by confining pulsations to an equatorial wave-guide
\citep{2003MNRAS.343..125T}, which is not
entirely visible to the observer. With the next generation of mode identification models promising the inclusion of more of these effects, it is hoped that this
will
lead to an explanation of some, or all, of the observed asymmetric phenomena.

\subsection{Spectroscopic and Photometric Frequencies}

Previous photometric and spectroscopic studies have identified frequencies and estimated the degree, $l$, of the corresponding modes. Analysis of the 2007
subset of the
spectroscopic line profile variations in \citet{2008AandA...489.1213U} found six frequencies, for which all but one was recovered in this extended
analysis. No evidence for the rotational frequency of $2.666$~\cd\ proposed by the above was found.
Two photometric frequencies in the same work were recovered but did not match those of the spectroscopy. The acquisition of high-temporal-resolution photometric
data from the CoRoT satellite allowed for the detection of $840$ significant photometric frequencies. The photometry and spectroscopy of this star
show very self-consistent results, even over a five-year time period, yet the two methods do not show the expected consistency between the observed
frequencies.  High-amplitude $\delta$ Scuti frequencies are weak or non-existent in the photometry and conversely, high-amplitude $\gamma$ Doradus frequencies are absent or have
low-amplitudes in the spectroscopic data. The particular absence of the strongest frequency from the photometry ($1.735$~\cd) in the spectroscopic results is
quite
surprising. Furthermore, three of the twelve frequencies that match between the photometry and spectroscopy were discarded as they had irregular phased residual
profiles. Although they are unlikely to be pulsational frequencies, their true origin is unknown.

There are some physical explanations for suppression of frequencies in photometric studies. The most obvious is the decreased sensitivity of photometric
methods to modes with high $l$, as the surface integrated flux changes weaken with each additional nodal line. This is also expected to affect spectroscopic
results, with smaller amplitude velocity variations across the star and increased noise in the standard deviation profiles. The low amplitude of the mode could be responsible for the lack of signal in the photometric data, but this is still a puzzling result.

Even more perplexing is the absence of the clearly dominant $f_{c1}$ in the spectroscopic results. There are no particular modes expected to be suppressed in
spectroscopic data but clear in photometry. There is the possibility that this frequency has another stellar origin, such as from a starspot. With an effective
temperature close to $7300$~K this star is close to the theoretical limit of convective blocking ($T > 7400$~K) above which spots are not predicted to occur.
The time range of the CoRoT run ($140$ days) does not exclude the possibility of spot frequencies arising in the data, but it would be coincidental to also
detect spots in the ground-based
photometry from \citet{2008AandA...489.1213U} spanning four years. It is not possible to exclude the possibility of $f_{c1}$ being a rotational or a
pulsational frequency, but, for either possibility, it is difficult to explain why it is not present in the spectroscopy.

\subsection{The Unresolved Hybrid Phenomenon}

The hybrid classification of HD\,49434 was proposed based on the two types of frequencies observed in the star. Inspection of the standard deviation profiles and
phased residual profiles of
the $\gamma$ Doradus group questions the independence of all but one of these frequencies. The remaining frequency, $f_{31}$, is of an extremely low amplitude. Other recent studies of hybrid stars have suggested most of the $\delta$ Scuti
frequencies to be non-intrinsic to the star, arising from interactions of the $\gamma$ Doradus frequencies with the strong $\delta$ Scuti or radial fundamental
modes \citep{2012AandA...540A.117C,2012arXiv1209.4836B}. A summary of the properties of the two hybrids compared with those of HD\,49434 is presented in Table
\ref{hybsum}. 

\begin{table*}\caption[Summary of the comparison of HD\,49434 to other hybrids.]{Summary of the
comparison of HD\,49434 to CoRoT~{105733033} \protect\citep{2012AandA...540A.117C} and \textsc{kic}~{8054146}
\protect\citep{2012arXiv1209.4836B}.}\label{hybsum}
 \begin{center}
\begin{tabular}{|lcccc|}
\toprule
&CoRoT~{105733033} & \multicolumn{2}{c}{HD\,49434 } & \textsc{kic}~{8054146}\\
&&Spec.& Phot. &\\
\midrule
Rotation & low $v_{rot}$  & \multicolumn{2}{c}{moderate \vsini\ } & high \vsini\ \\
 & obs. freq. splitting &\multicolumn{2}{c}{($87$~\kms)} & ($300$~\kms) high $v_{rot}$\\
\midrule
Strongest & $\gamma$ Dor & $\delta$ Sct & $\gamma$ Dor & $\gamma$ Dor\\
Frequencies & and radial & & & \\
\midrule
Phot. Amp.& 27 mmag & - & 0.134 mmag & 0.105 mmag\\
1st freq. & & & & \\
\midrule
Freq. Domains\footnotemark & yes & yes & no & yes\\
\midrule
Amp. Var. & not known & ? & not known & yes\\
 \midrule
Characteristic & frequency & no & no & no\\
 spacing & and periods & & \\
 \midrule
Other spacing & $F \pm f_{\gamma}$ & none & none & $f_{\delta} + f_{\gamma a}$ \& $f_{\delta}  - f_{\gamma b}$\\
\bottomrule
 \end{tabular}
 \end{center}
\footnotemark[1]\footnotesize{Is there an observed definition between the $\gamma$ Doradus and $\delta$ Scuti frequency regions?}
\end{table*}

The comparisons show that, although there is some suggestion that the CoRoT and \textit{Kepler} stars share some similar properties, HD\,49434 does not appear to
be
similar to
either. The high rotational velocity of \textsc{kic}~{8054146} may explain why no characteristic spacings are observed, yet the clarity of the
triplet and
``picket fence'' patterns show other regular behaviours. A low inclination, and thus a very fast rotation of HD\,49434, would explain some of the effects, such as
finding no gap between the frequency domains \citep{2012ASPC..462..111H} and could even explain some of the difference between the photometric and spectroscopic
frequency spectra.

The hybrid status of HD\,49434 is questioned in \citet{2009AIPC.1170..477B} since low, $\gamma$~Doradus range frequencies could be a result of
rotational splitting of the numerous high-degree p-modes. However, theoretical studies have shown that hundreds of frequencies are possible from an energy
viewpoint
\citep{2010ApJ...710L...7M}. Attempts to explain the origins of the modes through stochastic excitation have thus far been inconclusive
\citep{2010arXiv1003.4427C}. 

From the frequency analysis it appears that HD\,49434 is a pulsating star with multiple modes sharing ($l,m$) in the traditional $\delta$ Scuti frequency region and several skewed but possibly related modes sharing the same ($l,m$) in the $\gamma$~Doradus frequency region. In addition, there is a clear frequency in the $\gamma$~Doradus frequency region that appears to be unrelated to the others. This means HD\,49434 remains a candidate hybrid star. However this classification requires more evidence and understanding of properties of the line profile, rotational effects and hybrid nature before it can be confirmed.

\subsection{Summary}

Overall, the more data that are collected on HD\,49434, the more questions seem to be raised. The spectroscopic frequency results are surprising and show several
unexplained
phenomena, yet the clear distinction in the shape of the standard deviation and phase profiles of the higher and lower frequencies that are found suggests an
interesting physical origin. With the
recent studies of hybrids in the \textit{Kepler} field as a whole \citep{2011AandA...534A.125U,2011MNRAS.415.3531B,2010AN....331..989G} and individual CoRoT and
\textit{Kepler} hybrid stars \citep{2012AandA...540A.117C,2012arXiv1209.4836B} more questions are raised about these stars. The existence of the hybrid
phenomenon is
already in question and the spectroscopic results for HD\,49434 so far add weight to both sides of the debate. On the one hand most of the $\gamma$ Doradus frequencies
appear to be manifestations of the dominant $\delta$ Scuti mode. Yet the existence of just one $\gamma$ Doradus frequency that is visually distinct prevents the
classification of this star as a pure $\delta$ Scuti.

To determine the true nature of hybrids, it is clear that classification cannot be made based on frequencies alone until more is understood about the
physical origins of all the detected frequencies. Unfortunately, the alternative, collecting large numbers of high-resolution spectra, is telescope-intensive
and is limited to relatively bright stars. Observations at \textsc{mjuo} are limited to about seventh
magnitude stars, beyond which,
exposure times smear out the temporal resolution required to study $\gamma$ Doradus-type pulsations. This renders follow-up spectroscopy on most of the
hybrid stars discovered with CoRoT and \textit{Kepler} impossible at this time.

More observational data are always useful in the study of pulsations, but it is not evident that more data alone will answer the numerous questions arising from
HD\,49434. It appears to be a lack of understanding and modelling of the rotational aspects of the star that is hindering the mode identification the most.
More sophisticated models, and perhaps a new approach to the rotation problem, are required before further significant advances in the understanding of this
star can be made.

\section{Acknowledgements}
This work was supported by the Marsden Fund administered by the Royal Society of New Zealand.

The authors acknowledge the assistance of staff at
Mt John University Observatory, a research station of the University of
Canterbury.

We appreciate the time allocated at other facilities for multi-site campaigns and the numerous observers who make acquisition of large datasets possible.

This research has made use
of the {\sevensize SIMBAD} astronomical database operated at the CDS in
Strasbourg, France.

Mode identification results obtained with the software package {\sevensize FAMIAS} developed
in the framework of the FP6 European Coordination action {\sevensize HELAS}
(http://www.helas-eu.org/).

\label{lastpage}
\bibliography{references}{}
\bibliographystyle{mn2e}
\end{document}